\documentclass[longbibliography, aps, prb, preprint]{revtex4-2}
\usepackage[utf8]{inputenc}
\usepackage[version=4]{mhchem}
\usepackage{color}
\usepackage{hyperref}
\usepackage{graphicx}
\usepackage{setspace}
\usepackage{multirow}
\usepackage{braket}

\begin{document}
\title{Small hole polarons in yellow phase $\delta$-\ce{CsPbI3}}
\author{Yun Liu}
\email{liu\_yun@ihpc.a-star.edu.sg}
\affiliation{Institute of High Performance Computing (IHPC), Agency for Science, Technology and Research (A*STAR), 1 Fusionopolis Way, \#16-16 Connexis, Singapore 138632, Republic of Singapore}

\begin{abstract}
A heterophase containing both the optically active $\alpha$-\ce{CsPbI3} and non-active $\delta$-\ce{CsPbI3} has been demonstrated as an efficient white light emitter.
This has challenged the conventional perspective that non-active phases of perovskites are undesirable in any perovskite-based optoelectronic devices.
To understand the role that yellow phase $\delta$-\ce{CsPbI3} plays in the light emission process, we performed a systematic computational study on its electronic and optical properties, which are relatively unexplored in the literature. 
Using the Fr\"{o}hlich model we showed that both the electron and hole exhibit moderate coupling to longitudinal optical phonons.
Explicit density functional theory calculations show that small hole polarons exist with a formation energy of $-96$\,meV, corresponding to the contraction of the Pb-I bonds within a \ce{[PbI6]} octahedron.
Nudged elastic band calculations show that the hole polaron can hop into neighboring \ce{[PbI6]} octahedral sites with a small activation barrier of $2.1$\,meV.
Molecular dynamics simulations also show that the hole polaron exhibit periodic localization and delocalization behavior similar to carrier hopping with a characteristic lifetime of $0.3$\,ps. 
Our results have elucidated the role that $\delta$-\ce{CsPbI3} play in the self-trapped emission in perovskite-based white light emitting diodes by supporting the presence of a localized small hole polaron.
\end{abstract}
\maketitle

Lead halide perovskites have been widely investigated due to their desirable bandgaps, high photoluminescence quantum yields, low nonradiative rate, and solution processable synthesis\cite{stranks_metal-halide_2015, jena_halide_2019, liu_metal_2020, yoo_efficient_2021}.
The high optoelectronic performance of these perovskites-based devices rely on their crystal structures staying in the optically active black phases at ambient conditions.
During normal device operations, methylammonium lead iodides (\ce{MAPbI3}) are prone to chemical degradation upon exposure to moisture and oxygen, leading to the formation of undesirable side products such as \ce{PbI2}\cite{https://doi.org/10.1002/aenm.201904054, doi:10.1021/acs.chemrev.8b00336, domanski_systematic_2018, mosconi_ab_2015}.
Formamidinium (FA) and Cs-based compositions have improved chemical stability compared to their MA counterpart, but their optically active $\alpha$, $\beta$ or $\gamma$-phases can also quickly transform into the thermodynamically more stable nonperovskite hexagonal and orthorhombic $\delta$-phases respectively\cite{doi:10.1021/acsenergylett.0c00801}.
As these $\delta$-phases are optically inactive, they are typically detrimental to device performances.
Therefore, various strategies such as encapsulation, surface passivation and chemical alloying have been developed to improve their phase stability\cite{doi:10.1021/acsami.9b12579, doi:10.1021/acs.chemrev.8b00336, domanski_systematic_2018, yoo_efficient_2021, park_controlled_2023}.

While conventional wisdom dictates that the presence of yellow phases need to be minimized whenever possible, recent evidence suggests that they might play an important role in many interesting light emission phenomena.
For example, intrinsic quantum confinement with the co-existence of $\alpha$ and $\delta$-\ce{FAPbI3} has manifested as oscillatory absorption spectra\cite{wright_intrinsic_2020}.
It was hypothesized that the $\delta$-phases, with its larger bandgap, act as potential barriers resulting in nanostructures of $\alpha$-phases with a typical length scale of $10-20$\,nm.
The transformation into $\delta$-\ce{FAPbI3} can also be suppressed by alloying Cs atoms, leading to the ability to control the degree of quantum confinement\cite{doi:10.1021/acsnano.2c02970}.
Light emission from these quantum confined regions can be potentially used as single-photon sources analogous to colloidal quantum dots\cite{utzat_coherent_2019, kaplan_hongoumandel_2023}.

In addition, efficient and bright white light emitting diodes (LED) have been synthesized based on the heterophases of $\alpha$ and $\delta$-\ce{CsPbI3}\cite{chen_efficient_2021, doi:10.1021/acsnano.1c06849}. 
The carrier injection and transport were dominated by the optically active $\alpha$-phase, with the charges ultimately injected into the $\delta$-phase where efficient radiative recombination occurred \textit{via} self-trapped excitons (STE).
The synergetic cooperation between the two phases are hypothesized to achieve the high efficiency and brightness observed in the white LED.
A more recent investigation has found that the heterophases are a mixture of $\gamma$/$\delta$-phases, whereby alloying Cl$^{-1}$ at the halide site has improved film quality and regulated white light emission \cite{chen_nanoscale_2025}.

Carrier self-trapping in semiconductors occurs when excited carriers and the locally distorted lattice form a quasi-particle called polaron\cite{franchini_polarons_2021}.
Polarons can be classified into two types based on the strength of the electron-phonon coupling. 
Large polarons are formed due to the long-range interaction between electrons/holes and the longitudinal optical (LO) phonons. 
In halide perovskites, large polarons are known to be responsible for the long carrier diffusion lengths\cite{xing_long-range_2013, cinquanta_ultrafast_2019, ghosh_polarons_2020, meggiolaro_polarons_2020} and enhanced effective masses\cite{baranowski_polaronic_2024}.
Their formation kinetics have been observed in pseudocubic \ce{CsPbBr3}, and first-principles calculation showed that the localized charge density associated with the hole polaron can extend over 8 unit cells in one direction\cite{miyata_large_2017}.

Stronger coupling between charge carriers and phonons leads to larger local lattice distortions that provide the driving force for small polarons to form. 
The motion of small polarons have been observed in low dimensional perovskites and are characterized by incoherent phonon assisted hopping with much smaller carrier mobility\cite{smith_structural_2017, decrescent_even-parity_2020, han_exciton_2022,tao_momentarily_2021}.
The existence and formation mechanism of small polarons in 3D perovskites are still not conclusive\cite{fu_carriers_2023}.
For $\delta$-\ce{CsPbI3}, even the most basic optoelectronic properties are not understood, and it is therefore not clear what role it plays in the light emission processes in the heterophase devices.
Here in this work, we present a systematic investigation of the electronic structure of $\delta$-\ce{CsPbI3}. 
We utilized a combination of Fr\"{o}hlich polaron model, density functional theory (DFT) and \textit{ab initio} molecular dynamics (MD) calculations to show that small hole polarons are stable in $\delta$-\ce{CsPbI3}, which can potentially support carrier self-trapping and STE-mediated emissions.

\section*{Results and Discussion}
\begin{figure}[t]
 \includegraphics[scale=0.4]{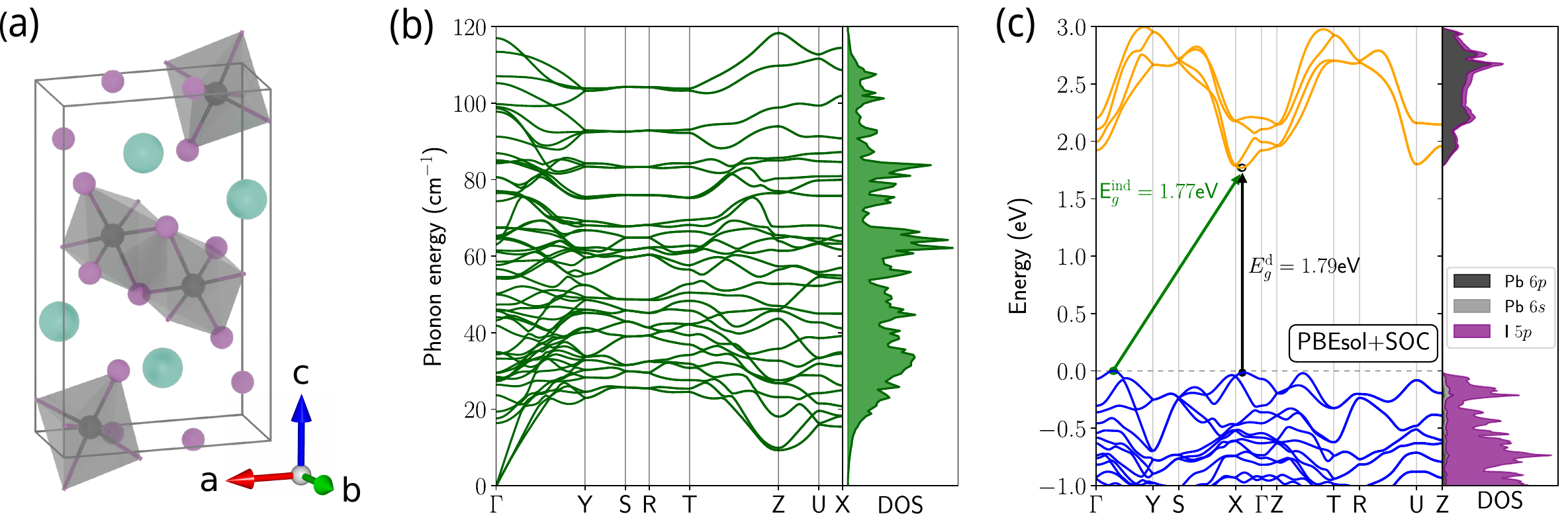}
 \centering
 \caption{Ground state properties. (a) Crystal structure of $\delta$-\ce{CsPbI3} where Cs, Pb and I atoms are represented as cyan, black and purple spheres with the \ce{[PbI6]} polyhedra drawn. (b) The phonon dispersion along high symmetry lines of the Brillouin zone with the total phonon density of states on the right panel. (c) Electronic band structure and projected density of states (PDOS) along high-symmetry lines at the PBEsol+SOC level, with the location of the indirect band edge locations ($E_g^{\mathrm{ind}}$) indicated by green circles, and that of the direct band edge locations ($E_g^{\mathrm{d}}$) indicated by black circles. High symmetry points within the Brillouin zone are defined as $\Gamma$=(0, 0, 0), Y=(0, 0.5, 0), S=(0.5, 0.5, 0), X=(0.5, 0, 0), R=(0.5, 0.5, 0.5), Z=(0, 0, 0.5), T=(0, 0.5, 0.5), U=(0.5, 0, 0.5).}
 \label{fig_band}
\end{figure}

We first evaluated the ground state electronic and structural properties of $\delta$-\ce{CsPbI3}, which belongs to the orthorhombic space group \textit{Pmna} with $20$ atoms per unit cell. 
Adjacent \ce{[PbI6]} octahedra share a common edge (Figure \ref{fig_band}(a)), in contrast to the corner sharing in $\alpha$, $\beta$ and $\gamma$-phases. 
The lattice parameters obtained with the PBEsol functional at $a=10.41$\,\AA\, $b=4.76$\,\AA, and $c=17.66$\,\AA\, agree well with the experimental values of $a=10.45$\,\AA\, $b=4.80$\,\AA,  and $c=17.76$\,\AA \cite{doi:10.1021/jacs.9b06055}. 
The detailed inequivalent Wyckoff positions and atomic coordinates are listed in Table S1 in the Supplementary Information.
The computed phonon dispersion and phonon density of states (DOS) are shown in Figure \ref{fig_band}(b), which exhibits no soft modes indicating dynamical stability.
For zone-center phonons, there are 60 modes that respect the $D_{2h}$ point group symmetry:
\begin{eqnarray}
    \Gamma = 3\Gamma_{\mathrm{acoustic}} + 30\Gamma_{\mathrm{Raman}} + 22\Gamma_{\mathrm{IR}} + 5 \Gamma_{\mathrm{silent}}
\end{eqnarray}
where 3 are acoustic phonon modes ($\Gamma_{\mathrm{acoustic}} = B_{1u} + B_{2u} + B_{3u}$).
For the optical phonons, 30 are Raman active ($\Gamma_{\mathrm{Raman}} = 10A_{g} + 5B_{1g} + 10B_{2g} + 5B_{3g}$), 22 are infrared (IR) active ($\Gamma_{\mathrm{IR}} = 9B_{1u} + 4B_{2u} + 9B_{3u}$) and 5 are optically silent($\Gamma_{\mathrm{silent}} = 5A_{u}$)\cite{liu_first-principles_2014}.
Table S2 contains the phonon frequencies and irreducible representations of the $\Gamma$-point phonons.
We also estimated the strength of the typical LO phonons to the transverse optical (TO) phonons by estimating the Lyddane-Sachs-Teller relationship, using the static and high-frequency dielectric constants $\Omega^2_{\mathrm{LO}}$/$\Omega^2_{\mathrm{TO}}$ = $\varepsilon_{\mathrm{static}}$/$\varepsilon_{\infty}$ = 1.30.

Figure \ref{fig_band}(c) shows the band structure and projected DOS (PDOS) of $\delta$-\ce{CsPbI3} including the effects of spin-orbit coupling (SOC). 
The valence band maximum (VBM) consists of mainly I $5p$ orbitals and the conduction band minimum (CBM) is dominated by Pb-I bonds made up of Pb $6p$ and I $5p$ orbitals.
$\delta$-\ce{CsPbI3} is an indirect bandgap semiconductor, as the positions of the VBM and CBM are located away from $\Gamma$ at (0, 0.15, 0) and (0.39, 0, 0) respectively.  
At the PBEsol+SOC level, the indirect bandgap $E_g^{\mathrm{ind}}$ is $1.77$\,eV while the direct bandgap $E_g^{\mathrm{d}}=1.79$\,eV is only about $20$\,meV higher.
Due to this small difference, $\delta$-\ce{CsPbI3} can be treated as an effective direct bandgap semiconductor. 
We note that the bandgaps are underestimated due to the well-known errors in semilocal DFT functionals.

\begin{table}
	\begin{tabular}{ c| c c  |c c | cc|}
		\hline \hline
		& \multicolumn{2}{c|}{$m*$ ($m_e$)} & \multicolumn{2}{c|}{$\alpha$} & \multicolumn{2}{c|}{$r_f$ (\AA)} \\
		& $e^-$ & $h^+$ & $e^-$ & $h^+$ & $e^-$ & $h^+$ \\
		\hline
		average & 0.51 & 0.86 & 3.25 & 4.22 & 21.9 & 19.2 \\
		$x$ & 0.65 & 0.82 & 3.68 & 4.12 & 20.6 & 19.4 \\
		$y$ & 0.26 & 0.82 & 2.31 & 4.12 & 26.1 & 19.4 \\
		$z$ & 2.11 & 0.96 & 6.61 & 4.46 & 15.2 & 18.7 \\
		\hline \hline
		
	\end{tabular}
	\caption{The electron and hole effective masses, dimensionless Fr\"{o}hlich polaron constants and Schultz polaron radii of $\delta$-\ce{CsPbI3} at $300$\,K.}
	\label{table_frohlich}
\end{table}

We then estimated the strength of the electron-phonon coupling ($\alpha$) using the Fr\"{o}hlich model (Details in the Supplementary Information). 
Within this model, the electron-phonon interaction is dominated by the LO phonon modes, which gives an estimate of the potential deformation the charge carriers on the surrounding lattice\cite{frohlich_interaction_1952}. 
Table \ref{table_frohlich} shows that $\alpha$ for electrons exhibit strong anisotropy. 
This can be explained from the highly anisotropic crystal structure of $\delta$-\ce{CsPbI3} whereby the corner sharing \ce{[PbI6]} octahedra extend infinitely along the $y$ direction giving rise to more disperse band structures. 
The octahedra are separate by the Cs atoms along the $x$ and $z$ directions, resulting in more flat bands as the Cs orbitals do not directly contribute to band edge states.
We note that the effective masses of the holes and electrons were extrapolated using a parabolic fit around the VBM and CBM positions corresponding to the indirect bandgap $E_g^{\mathrm{ind}}$.

The average values of $\alpha$ ($3.25-4.22$) are in the intermediate coupling regime ($3<\alpha<6$), larger than the typical values for the black phases ($1.17-2.68$)\cite{buizza_polarons_2021}. 
The increase is mainly due to the larger effective masses in $\delta$-\ce{CsPbI3}, as the characteristic phonon frequencies and dielectric constants are similar between the different phases.
The Fr\"{o}hlich coupling constants of $\delta$-\ce{CsPbI3} are closer to those of 2D and double perovskites, which means that it has the potential to exhibit similar spectroscopic signatures such as broadband emission arising from strong carrier self-trapping\cite{buizza_polarons_2021}. 
We then estimated the Schultz polaron radius ($r_f$) at $300$\,K\cite{schultz_slow_1959}. 
Ir also exhibits strong anisotropy; along the $z$ direction, $r_f$ is comparable to the $c$ unit cell lattice parameter, whereas along $x$ and $y$ directions, $r_f$ spans over multiple unit cells.

We want to highlight the importance of dimensionality for localized carrier formation.
In 3D and 2D systems, the local lattice deformations arising from the strong coupling between carrier and phonons generate a potential barrier acting as a trap, and free carriers need to overcome this barrier to self-trap\cite{RASHBA20001, 10.1143_JPSJ.63.637, 10.1143_JPSJ.48.472}.
In 1D, the barrier is effectively zero, and free and excited carriers can co-exist.
While $\delta$-\ce{CsPbI3} is structurally a 3D system, the strong anisotropy in its electronic structures show that its electronic dimensionality is closer to that of a  pseudo-1D system, as the the bonding orbitals arising from the \ce{[PbI6]} octahedra form 1D chains along the $b$ crystallographic direction\cite{woo_inhomogeneous_2023}.
The pseudo-1D electronic structure might favor easier formation of the self-trapped carriers in $\delta$-\ce{CsPbI3} compared to its black phase counterpart.

As the Fr\"{o}hlich model treats the electron-phonon interaction as a low order perturbation in a harmonic lattice potential, it has its limitations to fully describe the polaronic behavior in perovskites due to their soft anharmonic lattice\cite{schilcher_significance_2021}.
For an alternative assessment, we performed direct first-principles DFT calculations using hybrid functionals, which are used to accurately describe wavefunction localization associated with polarons by minimizing the self-interaction errors commonly associated with semilocal functionals\cite{ouhbi_polaron_2021, janotti_vacancies_2014, spreafico_nature_2014}.
We first needed to determine the fraction of exact Fock exchange that fulfills the Koopmans' conditions\cite{PhysRevB.82.115121}. 
To do so we calculated the occupied and unoccupied single particle energy levels related to transition of unrelaxed iodine and lead vacancies, by varying the amount of Fock exchange in unscreened PBE0 hybrid functional\cite{perdew_rationale_1996, adamo_toward_1999}. 
Figure S1 shows the computed band edges and the energy levels for the V$_\mathrm{I}$(+/0) and V$_\mathrm{Pb}$(-1/-2) transitions including the effects of SOC and finite-size corrections\cite{PhysRevLett.102.016402,falletta2020finite, PhysRevB.89.195205}.
The crossing between the different charge states of the same defect corresponds to the value of exact exchange for which the Koopmans’ condition is satisfied. 
For both defects, this crossing occurs at $0.12$.
This is substantially lower than the value of $0.28$ obtained for $\alpha$-\ce{CsPbI3}\cite{bischoff_nonempirical_2019}. 
We attribute the large difference to the lower defect tolerance of $\delta$-\ce{CsPbI3} which contains a larger octahedral tilt associated with the smaller Pb-I-Pb bond lengths and the weak anitbonding characteristic of the CBM state\cite{huang_intrinsic_2018}. 
This further illustrates that the different polymorphs of \ce{CsPbI3}, especially between the perovskite and nonperovskite phases, exhibit large differences in their electronic structures.
At the Fock exchange value of $0.12$, the bandgap is $2.67$\,eV, which is in good agreement with the experimental value of $2.93$\,eV\cite{valastro_optical_2021}, and much improved compared to the bandgap obtained earlier at the PBEsol level.

\begin{figure}[ht]
	\includegraphics[scale=0.5]{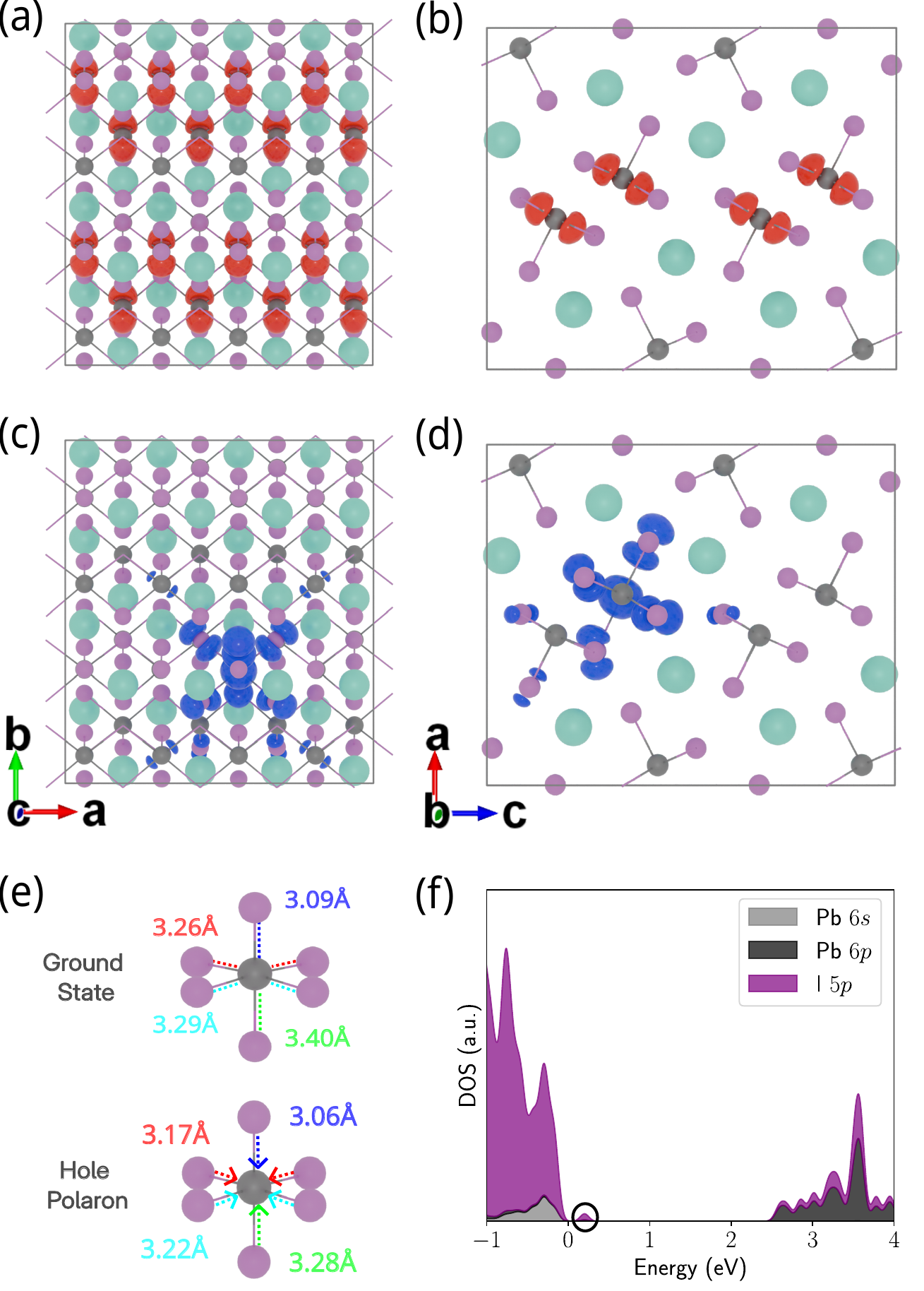}
	\centering
	\caption{Polarons in $\delta$-\ce{CsPbI3}. Wavefunction isosurface of the excess electron (a) and (b), and that of the hole polaron (c) and (d) as viewed from the $c$ and $b$ axes. (e) The Pb-I bond lengths of a \ce{[PbI6]} octahedron of the ground state structure and hole polaron with the arrows indicating the bond contractions. (f) PDOS of the relaxed polaronic structure, with the energetic location of the hole polaron circled.}
	\label{fig_polaron}
\end{figure}

We then applied the bond distortion method\cite{pham_efficient_2020, mosquera-lois_shakenbreak_2022, mosquera-lois_identifying_2023} to various bonds in $\delta$-\ce{CsPbI3} to break the crystal structure symmetries.
The atomic positions within the supercell were relaxed in the presence of an excess electron or hole to obtain the geometry of the localized polaron structure.
With an excess electron, the distorted structures always return to the ground state equilibrium(Fig. \ref{fig_polaron}[a][b]), with the wavefunction of the CBM state closely resembling that of the ground state(Fig. S2[a][b]). 
This shows that small electron polarons in $\delta$-\ce{CsPbI3} are unlikely to form.
For the excess hole, there is significant wavefunction localization around a single \ce{[PbI6]} octahedron(Fig. \ref{fig_polaron}(c)(d)), accompanied by the contraction of the Pb-I bonds by around 0.1\,\AA(Fig. \ref{fig_polaron}(e)). 
The hole polaron state appears just above the VBM (Fig. \ref{fig_polaron}(f)) with a formation energy of $-96$\,meV. 
Our DFT calculation shows that small hole polaron might be stable in $\delta$-\ce{CsPbI3}, and illustrates the limitation of the Fr\"{o}hlich model at characterizing systems with small polarons.

\begin{figure}[ht]
	\includegraphics[scale=0.5]{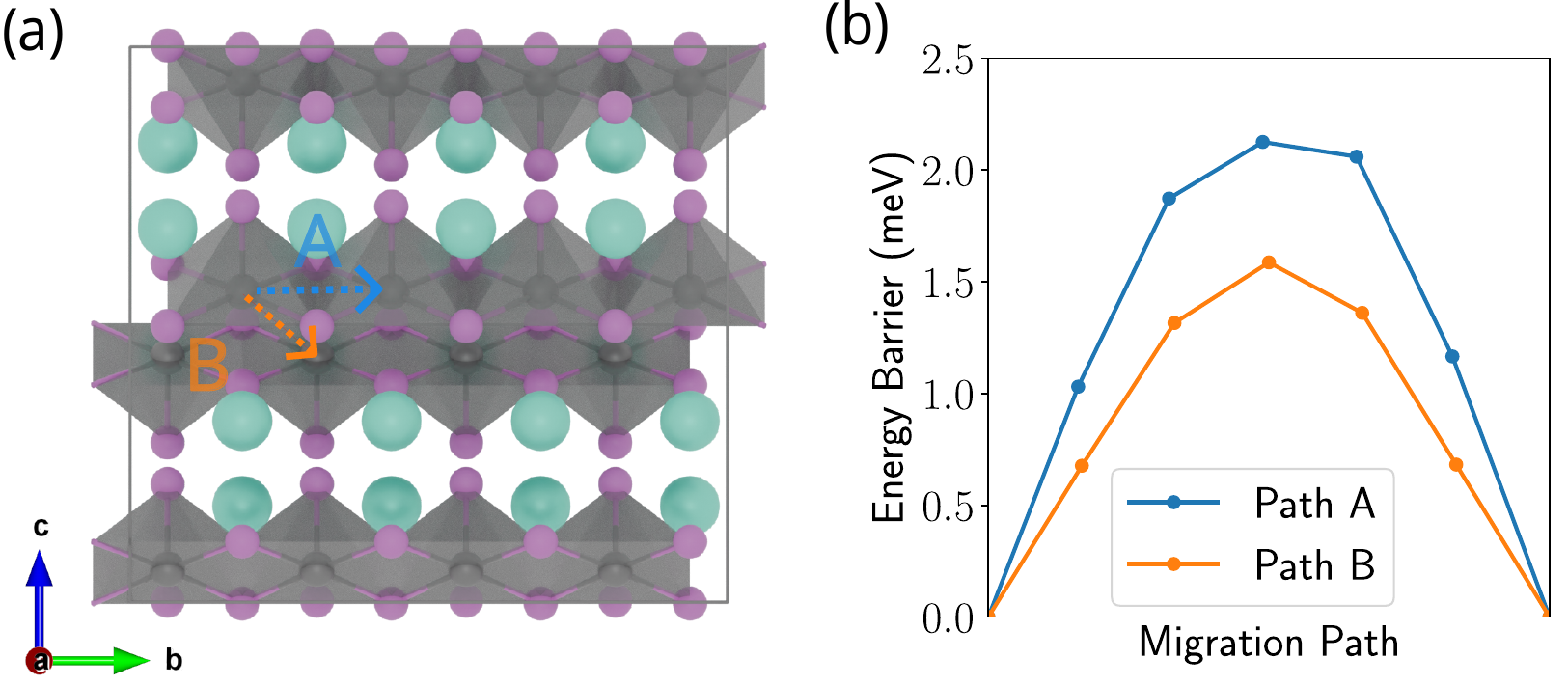}
	\centering
	\caption{Nudged elastic bands (NEB) calculation of hole polaron hopping in $\delta$-\ce{CsPbI3}. (a) The hole polaron migration pathways and (b) their corresponding energy barriers.}
	\label{fig_neb}
\end{figure}

The presence of small polarons can affect charge mobility in optoelectronic devices\cite{johannes_hole_2012, bjaalie_small_2015}. 
We therefore performed nudged elastic band (NEB) calculations to better understand its hopping behavior and obtain its activation barrier.
As the hole polaron is mainly localized onto a single \ce{[PbI6]} octahedron, it can hop in two general directions. 
One is along the crystallographic $b$ direction where the \ce{[PbI6]} octahedra are sharing edges along their square bipyramidal basal plane, labeled as path A in Fig. \ref{fig_neb}(a).
The hopping distance is approximately the distance between the neighboring Pb atoms at $4.89$\,\AA\, with an energy barrier of $2.1$\,meV (Fig. \ref{fig_neb}(b)).
The hole polaron can also hop along path B whereby the neighboring octahedra are sharing edges in one of their square pyramids. 
The hopping distance is approximately the same as path A at $4.81$\AA\, with a corresponding energy barrier of $1.5$\,meV.

Along both hopping paths, the hole polaron needs to migrate between two nearest neighbor I-I atoms.
The larger energetic barrier along path A is due to the smaller I-I distances at $4.40$\,\AA\, compared to that of $4.67$\,\AA\, along B.
However, hopping along path B would mean that the polaron need to change direction after each hop, so as to travel along the pseudo-1D \ce{[PbI6]} octahedral chain in the $b$ direction.
This means that for macroscopic transport, the polaron will hop along path A even though the energetic barrier is higher.
These energetic barriers are around 2 orders of magnitude smaller than some of the barriers observed in many oxides\cite{doi:10.1021/acs.jpcc.1c00702}.
The small hopping barrier also means that the presence of small localized hole polaron should not reduce the carrier mobility in 
$\delta$-\ce{CsPbI3}.

To further assess the possible presence of localized holes and electrons in the form of STE, we performed MD simulations of the system in a spin triplet configuration, which is typically more stable than the  spin singlet state\cite{PhysRevMaterials.7.085401}.
The electronic occupations are fixed using a constrained DFT approach (Details in Methods section).
The system was initialized using the distorted geometry of the hole polaron, and the degree of the electron and hole wavefunction localization was quantified using the Inverse Participation Ratio (IPR)
\begin{eqnarray}
    \mathrm{IPR} = \frac{\sum_i |\Phi_i|^4}{(\sum_i |\Phi_i|^2)^2} 
\end{eqnarray}
where $\Phi_i$ is the electronic wavefunction. 
In this definition of IPR, a localized state has a value of $1$.
For complete delocalization, the IPR tends towards $\frac{1}{V}=3.25\times 10^{-6}$ where $V$ is the volume or the total number of grid points used to sample the system. 

\begin{figure}[ht]
	\includegraphics[scale=0.75]{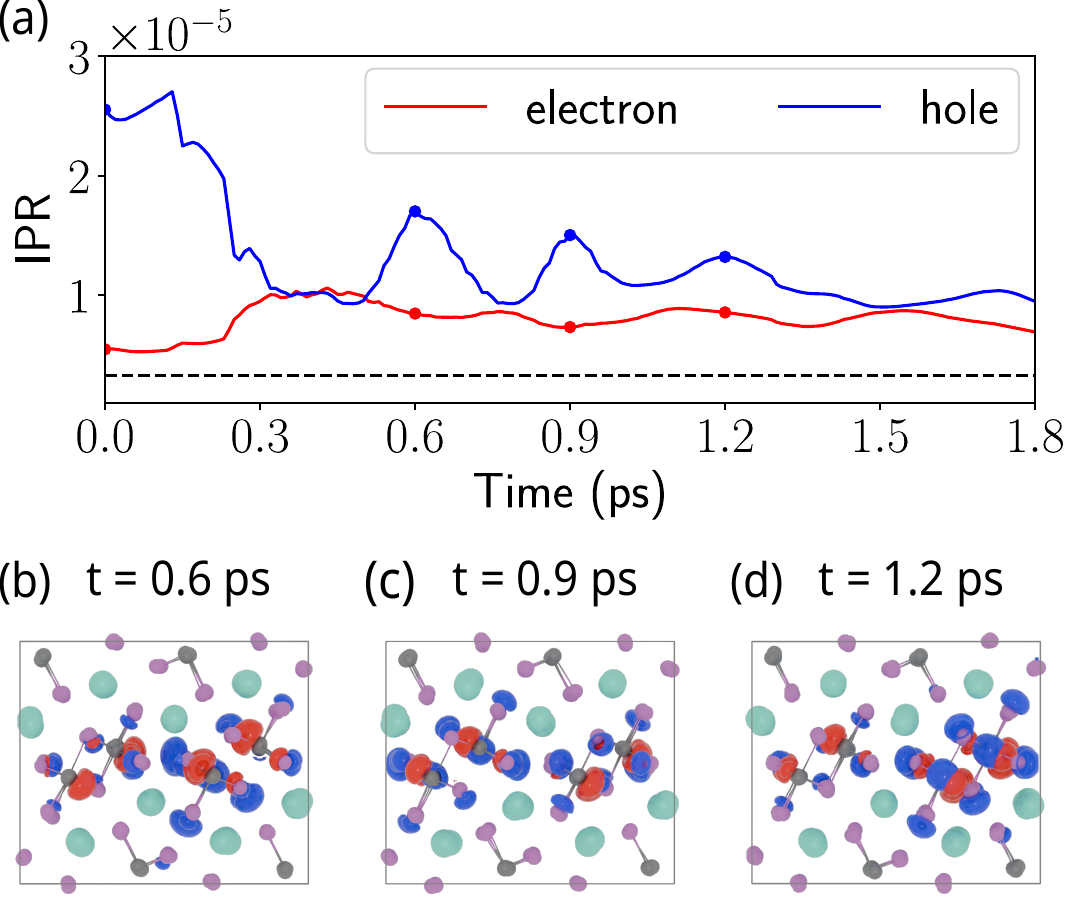}
	\centering
	\caption{Molecular dynamics (MD) simulation with the electron and hole in the spin triplet configuration. (a) The inverse participation ratio (IPR) of the electron and hole wavefunction during the MD run. The dashed line represents the IPR value of a completely delocalized system. The electron and hole wavefunction isosurface at (b) $t=0.6$\,ps (c) $t=0.9$\,ps and (d) $t=1.2$\,ps.}
	\label{fig_md}
\end{figure}

Fig. \ref{fig_md}(a) shows that for the duration of the MD run, the electron stays delocalized with the IPR value mostly staying below $1.0\times 10^{-5}$. 
On the other hand, the hole undergoes cycles of localization and delocalization corresponding to a hopping motion with a period of $0.3$\,ps between $0.6$ and $1.2$\,ps.
Fig. \ref{fig_md}(b)(c)(d) show the snapshots of the hole localization in different parts of the supercell at $0.6$, $0.9$ and $1.2$\,ps, respectively.
Overall, the MD results support conclusion from the static DFT calculations that small hole polarons can form in $\delta$-\ce{CsPbI3}, while small electron polaron is unlikely to form.

\section*{Conclusion}
In conclusion, we systematically investigated charge localization and carrier self-trapping in $\delta$-\ce{CsPbI3}. 
The Fr\"{o}hlich model predicts moderate coupling between the LO phonons and electrons and holes with the Fr\"{o}hlich polaron constants of around $3.25-4.22$ and Schultz radii of $\sim$ 20\,\AA\, at $300$\,K. 
Explicit DFT calculations using hybrid functionals and advanced structure searching methods show that small hole polaron is stable with a formation energy of $-96$\,meV. 
NEB calculations show that the energetic barrier for hopping is only $2.1$\,meV, much smaller than typical values associated with small polarons, and can facilitate high carrier mobility in $\delta$-\ce{CsPbI3} based optoelectronic devices.
MD simulations with the electron and holes in a triplet spin configuration further shows that the hole polaron exhibit characteristic hopping with a period of $0.3$\,ps.
Our results might have elucidated the role that $\delta$-\ce{CsPbI3} play in the self-trapped emission in perovskite-based white LED by supporting the presence of a localized small hole polaron.
There are some reports suggesting that electron polaron might be stable in $\alpha$-\ce{CsPbI3} clusters\cite{neukirch_polaron_2016}, highlighting the possibility that STE might exist in the heterophases of $\delta$-\ce{CsPbI3} and $\alpha$-\ce{CsPbI3}, with the electron polaron and hole polaron localized onto the different perovskite phases.
To fully understand the nature of self-trapped emission in perovskite based white LEDs, a systematic study of the structural and electronic properties of the heterophases is needed.

\section*{Methods}
Ground state electronic structure, phonon and polaron calculations were performed using the Vienna Ab initio Simulation Package (\textsc{VASP}, v6.4)\cite{kresse_efficient_1996, kresse_efficiency_1996}.
The core-valence interaction was described using the projector-augmented wave (PAW) method\cite{blochl_projector_1994}, with $9$ valence electrons for Cs ($5s^25p^66s^1$), $14$ valence electrons for Pb ($5d^{10}6s^26p^2$), and $7$ valence electrons for I ($5s^25p^5$).
The electronic wavefunctions were expanded in a plane wave basis with a cutoff of $400$\,eV.
For structural relaxations, atoms were relaxed until the Hellman-Feynman force converges below $10^{-3}$\,eV\AA$^{-1}$, and the volume until all components of the stress tensor are below $10^{-2}$\,GPa. 
The Brillouin zone of the unit cell was sampled with a $4\times8\times2$ $\Gamma$-centered Monkhorst-Pack\cite{monkhorst_special_1976} $\mathbf{k}$-point grid.
The band structure and projected density of states were computed using the PBEsol functional\cite{perdew_restoring_2008}, and spin-orbit coupling (SOC) effects are included due to the presence of heavy Pb atoms.
The Fr\"{o}hlich polaron properties were solved using the package \textsc{PolaronMobility}\cite{Frost2017}.

The phonon dispersion was computed using the finite displacement method with a $2\times4\times1$ supercell containing $160$ atoms with a $2\times2\times2$ $\mathbf{k}$-point grid as implemented in the \textsc{Phonopy} package\cite{togo_first_2015}. 
A non-analytical term was added to the dynamical matrix to treat the long range interaction arising from the macroscopic electric field induced by the polarization of collective ionic motions near $\Gamma$\cite{gonze_dynamical_1997}. 

For polaron and defects calculations, we used a $2\times4\times1$ supercell to minimize spurious interactions between periodic images, and sampled only the $\Gamma$-point. 
For the fulfillment of the Koopmans' condition, the corrections to the unoccupied Kohn-Sham eigenvalues of the defect-induced single particle levels were calculated as\cite{chen2013correspondence}
\begin{eqnarray}
   \epsilon_{\mathrm{corr}}^{\mathrm{KS}} = \frac{-2}{q}E_{\mathrm{corr}}
\end{eqnarray}
where $q$ is the charge of the defect, and $E_{\mathrm{corr}}$ is finite-size electrostatic correction. 
$E_{\mathrm{corr}}$ is computed using \textsc{sxdefectalign}\cite{PhysRevLett.102.016402}, where the diagonal terms of the static dielectric tensor ($\varepsilon_{\infty}$) are used for anisotropic screening.

After the addition or removal of an electron, structural relaxations were performed using spin-polarized calculations whereby the supercell lattice parameters are fixed and the atoms allowed to move, with the same force convergence criterion of $10^{-3}$\,eV\,\AA$^{-1}$. 
The binding energy of the hole polaron can be estimated by modifying the defect formation energy formula\cite{freysoldt_first-principles_2014}:
\begin{eqnarray}
  E_b = E_q[\mathrm{polaron}] - E[\mathrm{pristine}] + qE_\mathrm{VBM} + E_{\mathrm{corr}} 
\end{eqnarray}
where $E_q[\mathrm{polaron}]$ is the total energy of the distorted supercell of the hole polaronic state, $E$[pristine] is the total energy for the perfect crystal using an equivalent supercell, with $q$ denoting the excess of charge of a hole ($q=+1$), $E_{\mathrm{VBM}}$ is the position of the VBM. 
The diagonal components of the total dielectric tensor ($\varepsilon_0$) were used.

The MD simulation was carried out in the NVT ensemble in a $160$-atom supercell within \textsc{CP2K}\cite{vandevondele_quickstep_2005, doi:10.1063/5.0007045} code, with the lattice parameters set to the equilibrium volume obtained with hybrid functional (PBE0 with $0.12$ Fock exchange).
To model the self-trapped exciton (STE), the constrained DFT (cDFT) or $\Delta$SCF method was used to promote an electron from the VBM to the CBM in a spin triplet configurations with unrestricted spin-polarization.
We used DZVP-MOLOPT basis sets\cite{vandevondele_gaussian_2007, hartwigsen_relativistic_1998} and the plane wave cutoff energy of $400$\,Ry and core-valence interactions were described by Goedecker-Teter-Hutter pseudopotentials\cite{goedecker_separable_1996}.
The simulation was performed with the auxiliary density matrix method (ADMM)\cite{doi:10.1021/ct1002225}.
The temperature was set to $300$\,K and controlled by a Nosé-Hoover thermostat\cite{nose_unified_1984, hoover_canonical_1985} with a time step of $1$\,fs.

Crystal structures and isosurfaces are visualized using \textsc{Blender}\cite{blender} and \textsc{Beautiful Atoms}\cite{beautiful_atoms}.

\section*{Acknowledgements}
Y.L. acknowledges funding support from A*STAR under its Young Achiever Award and Career Development Fund (C233312001). This work is supported by A*STAR Computational Resource Centre (ACRC) and National Supercomputing Centre (NSCC), Singapore, through use of their high-performance computing facilities.

\bibliography{cspbi3}

\end{document}